\documentclass[sigconf]{acmart}

\AtBeginDocument{%
  \providecommand\BibTeX{{%
    \normalfont B\kern-0.5em{\scshape i\kern-0.25em b}\kern-0.8em\TeX}}}

\copyrightyear{2023}
\acmYear{2023}
\setcopyright{rightsretained}
\acmConference[FIRE 2023]{Forum for Information Retrieval Evaluation}{December 15--18, 2023}{Panjim, India}
\acmBooktitle{Forum for Information Retrieval Evaluation (FIRE 2023), December 15--18, 2023, Panjim, India}\acmDOI{10.1145/3632754.3632761}
\acmISBN{979-8-4007-1632-4/23/12}

\usepackage{pifont}
\newcommand{\cmark}{\ding{51}}%
\newcommand{\xmark}{\ding{55}}%

\usepackage{caption}
\usepackage{subcaption}
\begin{document}

\title[Expressivity-aware Music Performance Retrieval]{Expressivity-aware Music Performance Retrieval using \\ Mid-level Perceptual Features and Emotion Word Embeddings}



\author{Shreyan Chowdhury}
\affiliation{%
  \institution{Institute of Computational Perception}
  \city{Johannes Kepler University Linz}
  \country{Austria}}
\email{shreyan0311@gmail.com}

\author{Gerhard Widmer}
\affiliation{%
  \institution{Institute of Computational Perception}
  \city{Johannes Kepler University Linz}
  \country{Austria}}
\email{gerhard.widmer@jku.at}
\renewcommand{\shortauthors}{Chowdhury and Widmer}

\begin{abstract}
This paper explores a specific sub-task of cross-modal music retrieval. We consider the delicate task of retrieving a performance or rendition of a musical piece based on a description of its style, expressive character, or emotion from a set of different performances of the same piece. We observe that a general purpose cross-modal system trained to learn a common text-audio embedding space does not yield optimal results for this task. By introducing two changes -- one each to the text encoder and the audio encoder -- we demonstrate improved performance on a dataset of piano performances and associated free-text descriptions. On the text side, we use emotion-enriched word embeddings (EWE) and on the audio side, we extract mid-level perceptual features instead of generic audio embeddings. Our results highlight the effectiveness of mid-level perceptual features learnt from music and emotion enriched word embeddings learnt from emotion-labelled text in capturing musical expression in a cross-modal setting. Additionally, our interpretable mid-level features provide a route for introducing explainability in the retrieval and downstream recommendation processes.
\end{abstract}

\maketitle

\section{Introduction}
Music performance involves an interplay between a composer, a performer, and a listener. The performer imbues a composition with their own expression and style, and the listener perceives the emotion or mood being communicated through the music by the composer and the performer \cite{juslin2013does}. Thus the \textit{expressive quality} of performed music becomes an important attribute through which a musical performance is characterised \cite{Gabrielsson1996Emotional, akkermans2019decoding}. 
The present paper provides a step towards building music retrieval and recommendation systems that are sensitive to this expressive quality of music.

We aim to develop a system for retrieving a desired performance of a musical piece from a set of different performances based on a description of its style, expressive character, or emotion. This is an important problem when dealing with certain genres of music such as Western classical music, where we typically have several renditions of compositions performed by different artists\footnote{As stated by a major music streaming platform, \textit{``Classical music is different. It has longer and more detailed titles, multiple artists for each work, and hundreds of recordings of well-known pieces"} \cite{apple}.}, and possibly very distinguishing listeners. These interpretations, while all representing the same musical content (piece), could vary subtly or vastly in their expressive character, and traditional recommendation and retrieval algorithms are not designed to be sensitive to such performance characteristics. 

\begin{figure}[t]
  \includegraphics[width=\linewidth]{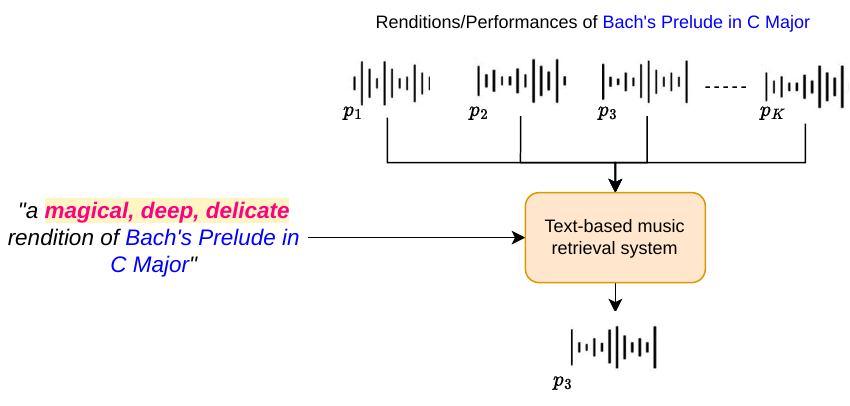}
  \caption{\small A system for retrieving the best-matching performance of a musical piece based on a text description of its expressive character.}
  \label{fig:aim}
\end{figure}

In our work, we look at text-based music retrieval. Formally, given a set of $K$ performances $P = \{p_1, p_2, \ldots, p_K\}$ of a musical piece and a query $Q = \{w_1, w_2, \ldots, w_N\}$ comprised of $N$ ``descriptive" words, we would like to retrieve the performance (say $p_3$, as shown in Figure~\ref{fig:aim}) that we believe a general audience might consider matches best with the description. This is a highly subjective task. We thus evaluate our models on the \textit{Con Espressione} dataset \cite{CancinoChacon2020OnTC}, which is a dataset comprised of different piano performances, by famous pianists, of a number of classical piano pieces, and associated free-text descriptions provided by contributors through an online questionnaire (more details in Section~\ref{sec:data}). In this dataset, we have on an average of five performances per piece, which thereby constitute the search space for each query. The assumption here is that we already know the piece (and thus the search space for that piece) for each query, and we want to retrieve the closest matching performance in terms of its expressive character. Inferring the piece itself from the query, while a necessary component of an end-to-end retrieval system, is not the focus of the present paper. The reader is encouraged to look at works on music search and retrieval such as \cite{schedl2014music} for more details on this topic.

The subjective nature of the task also motivated us to consider human-oriented, perceptually relevant descriptors for representing expressive qualities extracted from audio recordings. In particular, we focus on so-called \textit{Mid-level Perceptual Features} -- 
relatively high-level musical qualities that are considered to be perceptually important \cite{Aljanaki2018Midlevel}; in our work, these are learnt from human annotations.
Previous research has shown these features to have the capacity to predict musical emotions as well as to disentangle different performances based on emotion \cite{chowdhury2021perceived, chowdhury2022thesis, CancinoChacon2020OnTC}.

Furthermore, on the text side, we note the inefficiency of traditional word embeddings (Word2Vec, GloVe) to capture the intended emotion in descriptions of musical performances. Such language representation models carry the inductive bias that words used in the same context tend to possess similar meanings, thus resulting in emotionally dissimilar words like \textit{happy} and \textit{sad} having close proximity in the representation space, due to them often occurring in similar contexts in the training data. To derive correct emotional meanings from our text queries, we experiment with emotion-enriched word embeddings from Agrawal et al. \cite{agrawal2018learning} and find that they improve retrieval results significantly, particularly when combined with mid-level features on the audio side. 



\section{Related Work}\label{sec:related}
Our work sits in the broader area of cross-modal retrieval.
This is an area of active research in general, and is of great significance to music information retrieval \cite{muller2018cross, doh2023toward, li2019query}. Language-based audio retrieval is currently witnessing much interest from the research community as can be seen from active participation in competitions like DCASE \cite{dcase}. In this section, we look at some recent work in text-based retrieval for music.

\textbf{Audio-Language Learning for Music}: Text-based audio retrieval is typically done by learning a common embedding space of aligned audio and text embeddings. In ``MusCALL" \cite{Manco2022ContrastiveAL}, this is done using a multimodal contrastive learning (MCL) approach. Two encoders, $f_a(\cdot)$ for audio and $f_t(\cdot)$ for text, are learnt such that for any audio-text pair $(a_i, t_i)$, the resulting embeddings $z_{a,i} = f_a(a_i)$ and $z_{t,i} = f_t(t_i)$ lie close in the joint embedding space \cite{radford2021learning}. A ResNet-50 \cite{He2015DeepRL} is used for the audio backbone, and a downsized-Transformer \cite{Vaswani2017AttentionIA} for the text backbone. They use 250k audio-text pairs from a production music library as the dataset. In ``MuLan" \cite{Huang2022MuLanAJ} a much larger dataset of 44m recordings and weakly-associated, free-form text annotations is used. They use a pre-trained BERT \cite{devlin2018bert} as their text encoder and experiment with two different audio encoders -- ResNet-50 and Audio Spectrogram Transformer \cite{Gong2021ASTAS}. 

A more recent work is ``Music and Text Representation" learning (MTR) \cite{doh2023toward}. It lays out an investigation into effective design choices for universal representation learning for text-to-music retrieval systems. This work uses a set of 500k music-text pairs. They use a modified version of the Music Tagging Transformer \cite{won2021semi} as the audio encoder, and two different text encoders -- pre-trained GloVe \cite{pennington2014glove} and pre-trained BERT. We were able to obtain the trained models for this work, which we use as a baseline for our work.

\textbf{Emotion Embedding Spaces}: In \cite{Won2021EmotionES}, the authors propose ``emotion embeddings" for retrieval of musical pieces that match the emotional characteristic of stories. They use a ResNet model \cite{Won2020EvaluationOC} as the audio backbone and a pre-trained DistilBERT \cite{distilbert} as the text backbone. However, they do not use natural language in the audio-text pairs in their experiments; rather the text component comes from the labels or tags associated with the audio clips, which are mapped to the embedding space using a Word2Vec model or emotion lexicons. 

\subsection{Music-Text Representation (MTR)}
We consider the Music-Text Representation (MTR) model by Doh et al. \cite{doh2023toward} as the baseline system. It is a cross-modal retrieval system that projects audio and text representations onto a common embedding space and reduces the distance between paired vectors during training. It is trained on 500k audio-text pairs with the text formed by concatenating tags from different sources. This data is a subset of the Million Song Dataset \cite{bertin2011million} and contains songs from a mix of different genres. We use weights of the best version of their system according to the results on their paper, which is a contrastive model type with a BERT text encoder and stochastically sampled text representations.


\section{Datasets}\label{sec:data}

\textbf{The \textit{Con Espressione} Dataset} \cite{CancinoChacon2020OnTC} consists of recordings of 9 piano pieces played by different famous pianists (making a total of 45 performances), and associated free-text descriptions for each performance, collected from a large number of listeners. The study participants were asked to describe the \textit{expressive character} of each performance. Typical characterisations in the dataset are adjectives like ``cold'', ``playful'', ``dynamic'', ``passionate'', but also more complex phrases such as ``controlled with speed", ``smooth tempo variation", ``emotional with dynamics" etc. In this work, we use the aggregated answers for each performance. That is, all words or phrases used by different participants for a performance are concatenated into a single text description of the performance.

\textbf{The Mid-level Perceptual Features Dataset} \cite{Aljanaki2018Midlevel} consists of 5k song snippets of 15 seconds each annotated according to seven mid-level descriptors: \textit{melodiousness}, \textit{articulation}, \textit{rhythm stability}, \textit{rhythm complexity}, \textit{dissonance}, \textit{tonal stability}, and \textit{modality} (or `\textit{minorness}'). The ratings for the seven mid-level perceptual features were collected through crowd-sourcing. We use a model trained on this dataset as our audio encoder, and we refer to this trained model as the ``mid-level model", or ``mid-level encoder" in the following paragraphs. We add \textit{onset density} as an additional feature according to \cite{chowdhury2023decoding} (this gives the best results in our case).

\textbf{Additional Datasets: MusicCaps and Pitchfork Track Reviews}. In order to effectively train a model, we require a substantial amount of audio-text paired data, and unfortunately, the Con Espressione dataset is insufficient in size for this particular task. To expand our dataset, we look towards two sources: the MusicCaps dataset, and a trusted and well-known music review website, Pitchfork (\url{www.pitchfork.com}).

The MusicCaps dataset \cite{agostinelli2023musiclm} consists of 5.5k music clips from diverse genres with paired text descriptions

\section{Approach}
We define performance retrieval as the task of retrieving a particular musical performance from a set of $K$ performances $P = \{p_1, p_2, \ldots, p_K\}$ of a musical piece such that the returned performance best matches the query $Q = \{w_1, w_2, \ldots, w_N\}$ comprised of $N$ words. In our case, the query is a description of the desired expressive character of a performance provided in the form of text. For each query, the piece is known, so the search space for the system is the set of performances of that piece. This is arguably a hard task; the retrieval system will have to be sensitive to very subtle musical differences. In our experiments, we will consider a retrieval result "correct" if the output of the system ranks the performance corresponding to the input text the highest. 

We take the basic framework of Music-Text Representation (MTR) by Doh et al. \cite{doh2023toward} and modify the text and audio encoders. We hypothesise that two kinds of modifications might improve the model's effectiveness for expressivity-aware performance retrieval. First we need an audio encoder trained to extract features that are more attentive to the expressive qualities in music. Mid-level perceptual features \cite{Aljanaki2018Midlevel} have a significant capacity to capture such musical qualities and hold good predictive power for music emotion \cite{chowdhury2022thesis}. We thus replace the audio encoder in MTR with a mid-level feature model. This pre-trained model takes audio spectrograms as inputs and outputs seven mid-level features (seven real-valued scalars). To this we add an eighth feature: \textit{onset density} \cite{chowdhury2023decoding}, intended to model the `perceptual speed' of a performance. The assumption here is that points close together in the space spanned by the mid-level features are similar in their expressive quality. Second, we investigate if the system works better with a text encoder trained using emotion labels. While the text encoder in MTR is a state-of-the-art BERT sentence model, in our task, the sentence structure has less of a consequence than obtaining accurate representations of the descriptive words. We thus experiment with an emotion-enriched word embedding model (EWE) \cite{agrawal2018learning}. 

\subsection{Mid-level Features Audio Encoder and Emotion Enriched Text Encoder}

\begin{figure}[t]
  \includegraphics[width=0.9\linewidth]{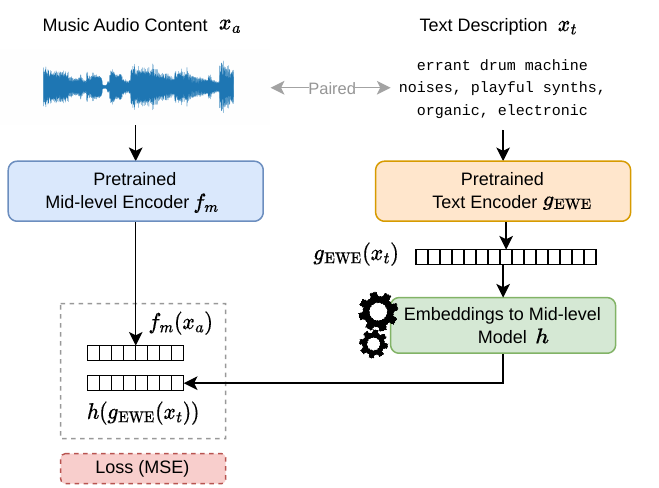}
  \caption{\small In our system, the audio and text encoders of a MTR model \cite{doh2023toward} are replaced by a mid-level feature model and emotion enriched word representation model respectively.}
  \label{fig:arch}
\end{figure}

As reasoned earlier in this paper, we propose replacing the audio encoder in MTR with a trained mid-level feature model ($f_m$ in Figure~\ref{fig:arch}). The input to this model is an audio spectrogram, the output is a vector of 8 mid-level features (including onset density). This model  is domain-adapted for piano music, since that is our domain of interest for this paper. As shown in \cite{chowdhury2021towards}, domain adaptation can improve mid-level prediction for piano audio without compromising the performance for other styles of music. 

The text encoder is also replaced by an emotion enriched word embedding (EWE) model ($g_{\text{EWE}}$ in Figure~\ref{fig:arch}), which outputs an embedding of dimension 300 for each word. For a set of descriptive words, we take the resultant vector by adding all individual embeddings element-wise. 

Now since these two encoders are pre-trained, they do not project to a common embedding space. Hence we need a model to project the outputs of the encoders to a common space for enabling cross-modal retrieval. We choose to project the text embeddings onto the mid-level feature space using a linear model $h$. Preserving the mid-level feature predictions has the additional advantage of providing explainable insight into the retrieval process, due to the dimensions possessing interpretable musical meanings. This is described in the next section. For $h$, a simple linear regression model proved sufficient. In some cases (see Table~\ref{tbl:aug}), transforming the text embeddings with principal component analysis (PCA) improved the results.

\subsection{Experiments and Results}
We perform piece-wise cross-validation over the Con Espressione dataset to fully utilise the available data, i.e., in each run the test set is the set of all performances of a piece in the dataset. The rest of the dataset is the train set for that run. This results in 9-fold cross-validation as we have 9 pieces in the dataset. For retrieval, we use the cosine similarity in the mid-level feature space:

\begin{equation}
\text{cosine\_similarity}(\mathbf{m_1}, \mathbf{m_2}) = \frac{\mathbf{m_1} \cdot \mathbf{m_2}}{\|\mathbf{m_1}\| \cdot \|\mathbf{m_2}\|}
\end{equation}

We use top-$k$ ratio and Mean Reciprocal Rank (MRR) as evaluation metrics. Top-$k$ ratio (ranges from 0 to 1, with higher scores indicating better retrieval performance) is defined as the number of queries for which the correct performance has a rank equal to or better than $k$ among all performances of the piece for which the query is made. We use $k=1$ and $k=2$. 

Mean Reciprocal Rank (MRR) is defined as the average of the reciprocal rank of the correct item retrieved over a set of queries:

\begin{equation}
\text{MRR} = \frac{1}{|M|}\sum_{i=1}^{|M|}\frac{1}{\text{rank}_i}
\end{equation}

where $|M|$ is the total number of queries, and $\text{rank}_i$ is the rank of the correct item in the result list for the $i$-th query. MRR ranges from 0 to 1, with higher scores indicating better performance of the ranking algorithm.

We also investigate the impact of augmenting the train set with additional data from the MusicCaps and Pitchfork datasets (see Table~\ref{tbl:aug}).

\begin{table}[t]
\small
\captionsetup{font=small}
  \caption{Retrieval results using different audio and text encoders. The values here are for a model trained with both Pitchfork and MusicCaps augmentation and no PCA on text embeddings (see Table~\ref{tbl:aug}). For the random baseline, performances are chosen at random. }
\centering
  \begin{tabular}{llrrr}
  \toprule
    
    Text Enc & Audio Enc & Top-1 & Top-2 & MRR  \\
    \midrule
    \multicolumn{2}{c}{(Random Baseline)}& 0.18 & 0.37 & 0.44\\
    MTR & MTR & 0.20 & 0.42 & 0.46  \\
    MTR & Mid-level & 0.22 & 0.53 & 0.50 \\
    EWE & MTR & 0.22 & 0.42 & 0.47 \\
    \textbf{EWE} & \textbf{Mid-level} & \textbf{0.38} & \textbf{0.64} & \textbf{0.61} \\
    
    \bottomrule
  \end{tabular}%
  \label{tbl:perf}
\end{table}

\begin{table}[b]
\small
\captionsetup{font=small}
  \caption{Effect of data augmentation and PCA on retrieval based on EWE embeddings and Mid-level features}
\centering
  \begin{tabular}{cc|c|rrr}
  \toprule
    \multicolumn{2}{c|}{Augmentation} & Text Emb & \multicolumn{3}{c}{Metrics} \\
    Pitchfork & MusicCaps & Transform (PCA) & Top-1 & Top-2 & MRR  \\
    \midrule

    \xmark & \xmark & \xmark & 0.22 & 0.46 & 0.48 \\
    \xmark & \xmark & \cmark & 0.29 & 0.49 & 0.52 \\
    \midrule
    \cmark & \xmark & \xmark & 0.24 & 0.44 & 0.48 \\
    \cmark & \xmark & \cmark & \textbf{0.40} & 0.58 & 0.60 \\
    \midrule
    \xmark & \cmark & \xmark & 0.33 & 0.57 & 0.57 \\
    \xmark & \cmark & \cmark & 0.27 & 0.53 & 0.52 \\
    \midrule
    \cmark & \cmark & \xmark & 0.38 & \textbf{0.64} & \textbf{0.61} \\
    \cmark & \cmark & \cmark & 0.36 & 0.56 & 0.57 \\
    
    \bottomrule
  \end{tabular}%
  \label{tbl:aug}
\end{table}

\begin{figure*}[t]
  \includegraphics[width=\textwidth]{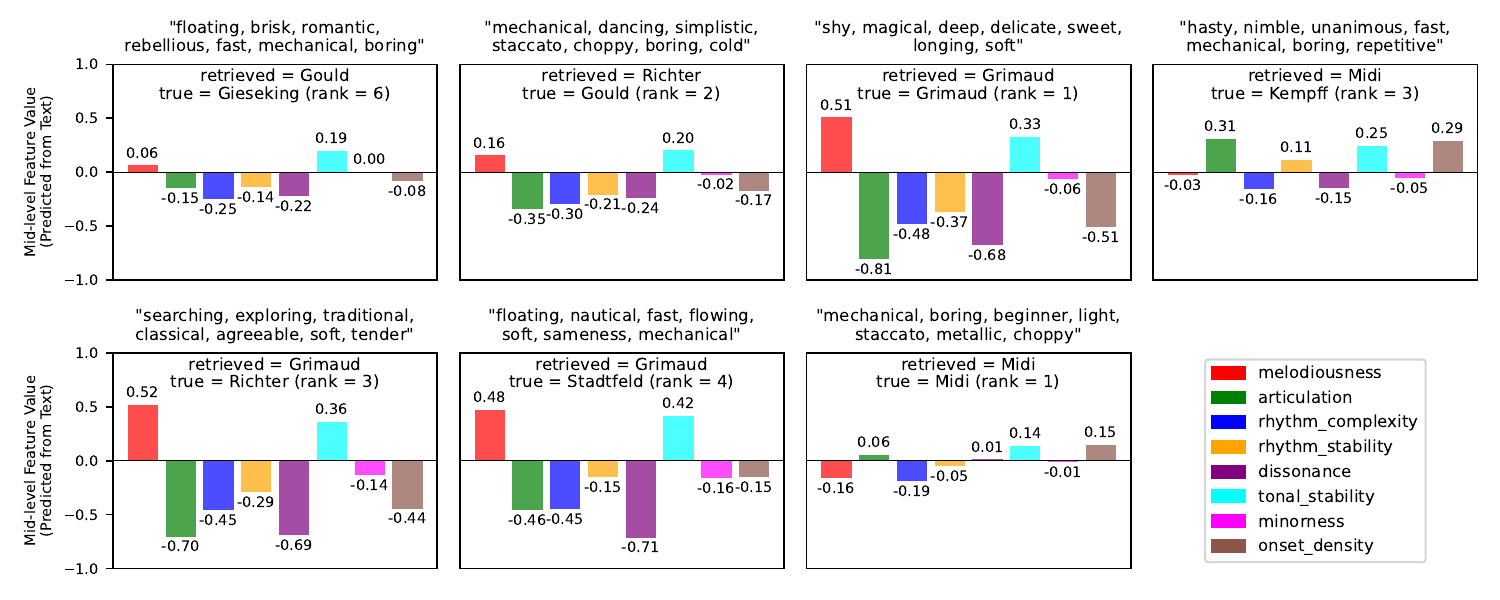}
  \caption{
  Seven performances of Bach's C Major Prelude, with human textual characterisation and corresponding mid-level feature values predicted by mapping $h(g_{\text{EWE}})$. Text and feature values relate to the performance identified by "true = ..."}
  \label{fig:bach}
\end{figure*}

Despite reasonable performance of traditional models on the general purpose task of music retrieval using text input, there remains much room for improvement for expressivity-aware music performance retrieval. We observe (see Table~\ref{tbl:perf}, second row) that for the MTR model the mean reciprocal rank for performance retrieval on the Con Espressione dataset is only 0.46, which means the average rank of the correct performance is greater than 2 (on sets of, on average, 5 performances per piece); in effect, this is similar to what we obtain with random guessing (Table~\ref{tbl:perf}, 1st row). Moreover, only 20\% of the queries return the correct performance as the top result.

On the other hand, our model with emotion word embeddings (EWE) and mid-level features returns the correct performance for 38\% of queries, with an MRR value of 0.61 (Table~\ref{tbl:perf}, last row), meaning the average rank of the correct performance is about 1.63. While our method shows significant improvement over the baseline, it is important to note that this is a highly subjective task and the dataset we have at hand is not large. The main objective of this paper is to provide a proof-of-concept that models trained using domain-specific perceptual features can lend a significant advantage in cross-modal retrieval applications.

From Table~\ref{tbl:aug}, we see that augmenting with additional data has a minor but positive effect on the results whereas PCA tends to improve performance for non-augmented datasets. For the augmented case, PCA actually worsened the performance. 



\section{Mid-level Features as Explanations}
It is instructive to look at an individual example. The mid-level features that our model predicts from the audio recordings, and which it uses to establish
a relationship to textual descriptions, can be viewed as a kind of \textit{explanation} of the model's retrieval choice, pointing to musical qualities in the performance that may have influenced the decision. In the example shown in Figure~\ref{fig:bach}, the piece in question is J.S.Bach's Praeludium in C major from his \textit{Well-Tempered Clavier} (Book I), for which our database contains seven different recordings, by Walter Gieseking, Glenn Gould, H\'el\`ene Grimaud, Wilhelm Kempff, Sviatoslav Richter, Martin Stadtfeld, and one completely expressionless, mechanical MIDI performance derived from the score. Given the query "shy, magical, deep, delicate, \ldots", the system returns the Grimaud performance (which is the "correct" one, as these descriptions were indeed associated with her performance by the human annotators); "mechanical, boring, beginner" returns, again correctly, the MIDI rendering. (The baseline MTR model also gets the Grimaud right, but returns Sviatoslav Richter in response to "mechanical, boring, beginner"). The mid-level feature profiles predicted from these text queries by our model (i.e., its `translation' of free text into mid-level feature space) explain its decisions quite well and conform both to our intuition, and to what we hear in the recordings. In particular, Grimaud is described as having high "melodiousness" and very low "articulation" whereas the MIDI rendering is characterised by rather neutral values throughout, and negative rather than positive "melodiousness". Such high-level musical descriptions could be useful as explanations in a real music search and recommendation application.

    

\section{Conclusion}
We present an experimental demonstration showing that mid-level perceptual features and emotion enriched text embeddings are useful in capturing some of the expressive musical character in audio recordings and relating them to descriptive text. We see significant improvements in all three metrics (particularly, the number of times the correct performance was retrieved almost doubled) and we note the effect of modifying \textit{both} the audio and the text encoders.
While this is not a large-scale study, our results point towards the importance of perceptually-driven features and emotion-aware models in music retrieval and consequently in music recommendation. Features that are interpretable and musically meaningful are also crucial for explainability and provide a route for training downstream interpretable models of music retrieval and recommendation.




\begin{acks}
This research was supported by the European Research Council (ERC) under the European Union's Horizon 2020 research and innovation programme, grant agreements No 670035 (``Con Espressione") and 101019375 (``Whither Music?").
\end{acks}

\bibliographystyle{ACM-Reference-Format}
\bibliography{biblio}

\end{document}